# Manipulating Berry curvature of $SrRuO_3$ thin films via epitaxial strain

Di Tian[1#], Zhiwei Liu[2,3#], Shengchun Shen[1#*], Zhuolu Li[1], Yu Zhou[1], Hongquan Liu[3,4], Hanghui Chen[3,4*] and Pu Yu[1,5,6*]

[1]*State Key Laboratory of Low Dimensional Quantum Physics and Department of Physics, Tsinghua University, Beijing, 100084, China*
[2]*Department of Electronics, East China Normal University, Shanghai 200241, China*
[3]*NYU-ECNU Institute of Physics, NYU Shanghai, Shanghai 200122, China*
[4]*Department of Physics, New York University, New York 10012, United States*
[5]*Collaborative Innovation Center of Quantum Matter, Beijing 100084, China*
[6]*RIKEN Center for Emergent Matter Science (CEMS), Wako 351-198, Japan*

Email: scshen@tsinghua.edu.cn; hanghui.chen@nyu.edu; yupu@tsinghua.edu.cn

**Author Contributions:** P.Y. designed the project. D.T and S.C. performed measurements and analyzed data. Z.L. and Y.Z. grew samples. Z.L. and H.C. performed first-principles calculations. H.C. and H.L. carried out model calculations. D.T., S.C, H.C. and P.Y. wrote the manuscript and all the authors commented on the paper.

**Competing Interest Statement:** The authors declare no competing interest.

**This PDF file includes:**

Main Text

Figures 1 to 4

**Berry curvature plays a crucial role in exotic electronic states of quantum materials, such as intrinsic anomalous Hall effect. As Berry curvature is highly sensitive to subtle changes of electronic band structures, it can be finely tuned via external stimulus. Here, we demonstrate in $SrRuO_3$ thin films that both the magnitude and sign of anomalous Hall resistivity can be effectively controlled with epitaxial strain. Our first-principles calculations reveal that epitaxial strain induces an additional crystal field splitting and changes the order of Ru $d$ orbital energies, which alters the Berry curvature and leads to the sign and magnitude change of anomalous Hall conductivity. Furthermore, we show that the rotation of Ru magnetic moment in real space of tensile strained sample can result in an exotic nonmonotonic change of anomalous Hall resistivity with the sweeping of magnetic field, resembling the topological Hall effect observed in non-coplanar spin systems. These findings not only deepen our understanding of anomalous Hall effect in $SrRuO_3$ systems, but also provide an effective tuning knob to manipulate Berry curvature and related physical properties in a wide range of quantum materials.**

## Significance

Berry phase and Berry curvature contribute significantly to a large collection of intriguing properties in quantum materials, particularly intrinsic anomalous Hall effect (AHE). One of the central topics is to explore the evolution of Berry curvature and related properties under external stimulus. Here we report the epitaxial strain effectively tunes the Berry curvature and the magnitude and sign of corresponding intrinsic AHE in $SrRuO_3$ films, which is due to modification of electronic structures induced by strain. In the same vein, we further reveal the Berry curvature can also be modulated with the rotation of magnetization, resulting in an exotic nonmonotonic change of anomalous Hall resistively. Our findings suggest that epitaxial strain provides a nice approach for manipulating Berry curvature and its related physics.

# Introduction

Over the past decades, Berry phase and Berry curvature have become an important ingredient in condensed matter physics, underlying a spectrum of phenomena such as ferroelectric polarization, orbital magnetism, and various Hall effects (1-5). In material systems that lack time-reversal symmetry or space-inversion symmetry or both, electrons acquire an anomalous velocity that is transverse to the external electric field and consequently gives rise to an anomalous Hall current, which contributes to intrinsic anomalous Hall effect (AHE) (6-10). Specifically, the intrinsic contribution of AHE is determined by the integration of Berry curvature over all the occupied electronic bands in the entire Brillouin zone. Therefore, its sign and magnitude highly depend on the topology of electronic band structure. Especially in a ferromagnetic material with sizable spin-orbit coupling (SOC), the Berry curvature can be substantially increased by an avoided band crossing near the Fermi level, leading to a considerable contribution to AHE (11-16). Thus, the evolution of intrinsic AHE represents a convenient probe to understand how the electronic structure changes Berry curvature under external stimulus.

Transition metal oxide $SrRuO_3$ provides an ideal platform to investigate Berry curvature, because it is a $4d$ metallic ferromagnet with a sizable SOC (17-20), in which the AHE mainly arises from the Berry curvature of its electronic band structure (11, 21, 22). Due to the strong coupling between lattice distortions and electronic structure in transition metal oxides, previous studies have already revealed that epitaxial strain has profound effects on magnetic properties of $SrRuO_3$ films (23-32). These results suggest that epitaxial strain might be an effective approach to manipulating Berry curvature and related AHE in $SrRuO_3$ thin films, which however has not been comprehensively investigated yet. Besides, motivated by recent observations of unconventional Hall effect with a pronounced hump feature in $SrRuO_3$ ultra-thin films and heterostructures (33-39), the underlying mechanism in which whether it belongs to trivial AHE caused by Berry curvature or exotic topological Hall effect due to non-coplanar spin textures is currently under intensive debate.

In this work, we carry out systemic transport studies and detailed first-principles calculations to investigate the AHE in both tensile- and compressive-strained SrRuO$_3$ pseudocubic (001)-orientated thin films. We find that 1) as the epitaxial strain changes from compressive to tensile, SrRuO$_3$ thin films exhibit a robust ferromagnetic state, but their magnetic easy axis evolves from out-of-plane direction toward in-plane direction. 2) The saturated anomalous Hall resistivity (or conductivity) of SrRuO$_3$ thin films changes both its sign and magnitude with epitaxial strain. Theoretical calculations reveal that epitaxial strain induces an additional crystal field splitting and changes the order of Ru $d$ orbital energies, which results in the sign and magnitude change of the anomalous Hall conductivity. 3) In SrRuO$_3$ thin films under tensile strain, rotation of Ru magnetic moments in real space with magnetic field affect the Berry curvature, which leads to a nonmonotonic change of the anomalous Hall resistivity at the intermediate field region. While this AHE signal resembles the exotic 'hump feature' observed in other systems, its underlying mechanism is directly correlated with the evolution of Berry curvature with spin rotation. These results altogether demonstrate that epitaxial strain is a powerful approach to tuning the underlying electronic state, Berry curvature and thus anomalous Hall effect in a wide range of quantum materials.

## Results

**Magnetic anisotropy modified by epitaxial strain**

Pseudocubic (001) oriented SrRuO$_3$ thin films were coherently grown on (LaAlO$_3$)$_{0.3}$(Sr$_2$TaAlO)$_{0.7}$ (LSAT) (001), SrTiO$_3$ (STO) (001), DyScO$_3$ (DSO) (110)$_o$, and GdScO$_3$ (GSO) (110)$_o$ substrates by pulsed laser deposition method (see the Supplemental Materials). Crystallinity and lattice parameters of SrRuO$_3$ films on different substrates were determined by high-resolution X-ray diffraction (XRD). Figure 1a shows the θ-2θ scans of SrRuO$_3$ films grown on various substrates, in which the clear pseudocubic (001) and (002) peaks for both SrRuO$_3$ and substrates confirm the pseudocubic *c* axis orientated film growth. All the films are fully strained and epitaxially grown on substrates, as examined by X-ray reciprocal space mapping (RSM), in which the peaks of SrRuO$_3$ and substrates are aligned on the same horizontal vectors (Fig. S1). Therefore, the corresponding in-plane strains are calculated to be -1.4%, -0.46%, +0.51%, and +0.97% for LSAT, STO, DSO and GSO, respectively. Figure 1b shows the temperature dependence of longitudinal resistivity $ρ_{xx}$ measured on all strained thin films, which exhibit metallic behavior down to lowest temperature and similar ferromagnetic transition temperature $T_c$ around 150 K as indicated by the kink features in $ρ_{xx}$ (*T*) curves (Fig.S2).

With the obtained high quality ferromagnetic metallic strained SrRuO$_3$ thin films, we proceed to investigate the corresponding change of magnetic anisotropy, as the strong SOC induced uniaxial anisotropy can be coupled with the epitaxial strain (19, 28). Figure 1c shows the out-of-plane (OOP) magnetoresistance (MR) measurements at 25 K for various strained SrRuO$_3$ films. For thin films under compressive strains (on STO and LSAT), a standard negative MR with butterfly-shape hysteresis is observed. While for tensile-strained films (on DSO and GSO), the OOP MR changes into a nonmonotonic field dependence with positive *MR* observed for field region between the coercive- and saturated-magnetic fields, which suggests the deviation of magnetic easy axis from OOP direction. This point is further clarified by the anisotropic magnetoresistance (AMR) measurements (Fig. 1d). The reversal of magnetic moment leads to peaks in $ρ_{xx}$ around easy axis when the angle between the easy axis and the field exceed 90°. For compressive strained thin films, the AMR results reveal peaks in $ρ_{xx}$ with hysteresis centered around θ = 90° (or 270°) in the clockwise and counter-clockwise field rotation, indicating the magnetic easy axis is close to OOP direction. In contrast, for thin films under tensile strain, the AMR shows hysteretic peaks with centered around $\theta$ = 180º, the magnetic easy axis is close to IP direction. This result is consistent with previous studies on strained SrRuO$_3$ films (28, 31).

To provide theoretical support for these experimental results, we perform *ab initio* calculations on the total energy of ferromagnetic SrRuO$_3$ with its magnetization pointing along the IP and OOP directions. Figure 1e shows the DFT-calculated energy difference between these two magnetic states of SrRuO$_3$ as a function of epitaxial strain. We find that when the applied strain changes from compressive to tensile, the magnetic easy axis of SrRuO$_3$ evolves from OOP to IP, which is qualitatively consistent with our experimental observations.

**AHE and Berry curvature tuned by epitaxial strain**

Knowing the fact that epitaxial strain can dramatically modify the magnetic anisotropy through the lattice-spin coupling, we turn to the investigation of AHE in strained SrRuO$_3$ thin films. Figure 2a shows the anomalous Hall resistivity $\rho_{xy}^{AHE}$ measured as a function of magnetic field at 25 K for thin films under different strains. For those films under compressive strain, a clear square-like hysteresis loop is observed, while for those films under tensile strain, the hysteresis feature is suppressed and we observe a non-monotonic behavior at the intermediate field region. We first focus on the anomalous Hall resistivity (or conductivity) under strong magnetic field, in which the magnetization of all SrRuO$_3$ thin films is fully saturated along the OOP direction (Fig. S3). The saturated $\rho_{xy}^{AHE}$ at low temperatures (taken at a fixed magnetic field of -9T) changes from positive for compressive-strained thin films to negative for tensile-strained thin films (Fig. 2b). Besides, for compressive-strained thin films, $\rho_{xy}^{AHE}$ demonstrates a nonmonotonic evolution with the temperature, which stays positive in a wide range of temperature (up to 125 K), and then undergoes a sign change to negative and eventually decreases to zero above $T_c$, which is consistent with the previous study of AHE in SrRuO3 (11). While $\rho_{xy}^{AHE}$ stays negative without any sign changes with increasing temperature for tensile-strained thin films.

Previous works suggest the AHE in SrRuO3 mainly arises from the Berry curvature of its electronic structure [11,27,28]. To further explore the modification of AHE of SrRuO3 by epitaxial strain, we perform first-principles calculations on the Berry curvature of SrRuO3 under different strains (Figs. S4 and S5). Figure 2c and Fig. S6 show the calculated intrinsic anomalous Hall conductivity $\sigma_{xy}^{AHE}$ $(= -\rho_{xy}^{AHE}/\rho_{xx}^2)$ that arises from the Berry curvature as a function of epitaxial strain. We find that as the strain changes from compressive to tensile, $\sigma_{xy}^{AHE}$ evolves from negative to positive, which is qualitatively consistent with our experimental results. Since the intrinsic anomalous Hall conductivity is the integral of negative Berry curvature over all the occupied states in the entire Brillouin zone, we explicitly present the negative Berry curvature $(-\Omega^z(\mathbf{k}))$ in the $(k_x, k_y)$ plane with $k_z = 0$ with different epitaxial strains (top: 2% compressive strain, middle: no strain and bottom: 3% tensile strain) in Fig. 2d. As the strain changes from compressive to tensile, the valleys in $(-\Omega^z(\mathbf{k}))$ at $k_z = 0$ plane are suppressed but some peaks develop along Γ to *Y* **k**-path. Specifically, under 2% compressive strain, $(-\Omega^z(\mathbf{k}))$ at $k_z = 0$ plane exhibits deep valleys along Γ to *Y* **k**-path, while under 3% tensile strain, $(-\Omega^z(\mathbf{k}))$ at $k_z = 0$ plane exhibits a number of peaks along Γ to *Y* **k**-path. These results demonstrate that Berry curvature can qualitatively account for the sign and magnitude change of $\rho_{xy}^{AHE}$ (or $\sigma_{xy}^{AHE} = -\rho_{xy}^{AHE}/\rho_{xx}^2$) induced by epitaxial strain in the above experimental observations.

**Electronic structure modified by epitaxial strain**

Next, we perform a more detailed analysis to understand how exactly epitaxial strain affects the underlying electronic structure of SrRuO$_3$, which subsequently changes its Berry curvature and anomalous Hall conductivity. Figures 3a, 3b, 3c show the band structure of SrRuO$_3$ along $\Gamma$ to $Y$ and the corresponding negative Berry curvature $(-\Omega^z(\mathbf{k}))$ under 2% compressive strain, no strain and 3% tensile strain, respectively. Consistent with previous studies (11, 12, 40), the valley in the $(-\Omega^z(\mathbf{k}))$ of compressively strained SrRuO$_3$ and the peak in the $(-\Omega^z(\mathbf{k}))$ of tensilely strained SrRuO$_3$ both arise from a near-Fermi-level avoided band crossing, whose small gap is opened by spin-orbit interaction. We note that the two near-degenerate states that are close to the avoided band crossing only contribute significantly to the Berry curvature when the Fermi level lies within the spin-orbit-induced gap. However, under both 2% compressive and 3% tensile strains, the Fermi level cuts through the lower band of the avoided band crossing and thus a small electron pocket is formed. This causes the sudden quench of the Berry curvature because in the electron pocket (highlighted by two vertical dashed lines) the two near-degenerate states are both occupied and therefore make negligible contribution to the Berry curvature.

We note that Berry curvature is a $k$-dependent property, while the anomalous Hall conductivity $\sigma_{xy}^{AHE}$ is a 'global' property that involves the integral of Berry curvature over all occupied bands in the entire Brillouin zone. To reveal how the electronic structure of SrRuO$_3$ changes the sign and magnitude of its anomalous Hall conductivity via epitaxial strain, we further carry out two 'thought experiments' to simulate two important effects on SrRuO$_3$ electronic structure from epitaxial strain. One is that epitaxial strain changes the oxygen octahedral rotations of SrRuO$_3$ (31, 41), which affects the band width of Ru $d$ orbitals. The other is that epitaxial strain changes Ru-Ru bond lengths and induces additional crystal field splitting between different Ru $d$ orbitals. The first thought experiment is to artificially 'turn off' the oxygen octahedral rotations in SrRuO$_3$ and re-calculate the anomalous Hall conductivity. Figure 3d shows that in the artificial SrRuO$_3$ crystal, which does not have oxygen octahedral rotations, $\sigma_{xy}^{AHE}$ also changes its sign (from negative to positive) as the epitaxial strain evolves from compressive to tensile. This trend is very similar to actual distorted SrRuO$_3$ with $a^-a^-c^+$ oxygen octahedral rotations (see Fig. 3c), which rules out oxygen octahedral rotations as the main reason to account for the sign change of $\sigma_{xy}^{AHE}$ and also greatly simplifies our modelling for $\sigma_{xy}^{AHE}$ of SrRuO$_3$. Hence, we build a simple tight-binding model for pseudo-cubic SrRuO$_3$ $H(k) = H_0(k) + H_{\text{soc}}$, where $H_0(k)$ describes SrRuO$_3$ without SOC and $H_{\text{soc}}$ describes the SOC effects on Ru atoms. The bases for $H(k)$ are five Ru *d* orbitals with two spins: $|z^2 \uparrow\rangle, |xz \uparrow\rangle, |yz \uparrow\rangle, |x^2 \uparrow\rangle, |xy \uparrow\rangle, |z^2 \downarrow\rangle, |xz \downarrow\rangle, |yz \downarrow\rangle, |x^2 \downarrow\rangle$ and $|xy \downarrow\rangle$ (42). Under these bases, $H_0(k)$ is a block diagonal matrix within each spin and can be obtained by using maximally localized Wannier functions. While for the spin-orbit coupling $H_{\text{soc}} = \lambda \vec{L} \cdot \vec{S}$, where $\lambda$ is the atomic spin-orbit-coupling constant, we find that $\lambda = 0.13$ meV is reasonable to describe the SOC effects of pseudo-cubic SrRuO$_3$ (see the Supplementary Materials for the details of this model). The advantage of this simple model is that we can artificially control the Ru orbital energy so as to mimic the additional crystal field splitting induced by epitaxial strain. In an exact pseudo-cubic SrRuO$_3$, three Ru *t$_{2g}$* orbitals are degenerate and two Ru *e$_g$* orbitals are degenerate. Due to the anti-bonding nature of Ru *d* state, under compressive (tensile) strain, the $d_{xy}$ orbital should have higher (lower) energy than $d_{xz}$ and $d_{yz}$; the $d_{x^2-y^2}$ orbital should have higher (lower) energy than $d_{3z^2-r^2}$. We can always add an energy shift on all orbitals, which is counteracted by a shift in the chemical potential (this does not change any physical properties). Figure 3e shows the schematics of the Ru *d* orbital energies in the exact cubic case versus under compressive/tensile strains. A 'global' energy shift is made on all Ru *d* orbitals, so that Ru $d_{3z^2-r^2}$ orbital has the same energy in all three cases. $\delta_1$, $\delta_2$, $\delta_3$ are the new energies of Ru $d_{xz}/d_{yz}$, $d_{xy}$ and $d_{x^2-y^2}$ orbitals relative to their values of the cubic structure (i.e. in the absence of strain), respectively. $|\delta_3|$ is the largest because Ru $d_{x^2-y^2}$ orbital forms a $\sigma$-bond with O $p$ orbitals, $|\delta_2|$ is the second largest because Ru $d_{xy}$ is an in-plane orbital and $|\delta_1|$ is the smallest because Ru $d_{xz}/d_{yz}$ are out-of-plane orbitals. To simplify our modelling and use only one control parameter, we set $\delta_2 = 2\delta_1$, $\delta_3 = 4\delta_1$, while we note that the qualitative result does not depend on this assumption (see the Supplementary Materials and Fig. S7). Then we calculate $\sigma_{xy}^{AHE}$ of SrRuO3 of the tight-binding model $H(k) = H_0(k) + H_{soc}$ as a function of $\delta_1$. Figure 3f shows that a positive $\delta_1$ mimics the effects from compressive strain and when $\delta_1$ is sufficiently large, $\sigma_{xy}^{AHE}$ becomes negative; a negative $\delta_1$ mimics the effects from tensile strain and $\sigma_{xy}^{AHE}$ stays positive. This result is qualitatively consistent with Fig. 3d in which epitaxial strain is imposed. Figure 3f clearly demonstrates that the additional crystal field splitting between Ru $d$ orbitals induced by epitaxial strain, is the key reason for the sign and magnitude change of $\sigma_{xy}^{AHE}$ in SrRuO3 thin films.

**Magnetization rotation induced Berry curvature evolution**

Finally, we turn our attention to the nonmonotonic hump features observed at the intermediate magnetic-field region in tensile-strained SrRuO3 thin films. Figure 4a shows the field dependent $\sigma_{xy}^{AHE}(= -\rho_{xy}^{AHE}/\rho_{xx}^2)$ of SrRuO3 grown on DSO substrate at a number of temperatures, where the hump features can be clearly visualized. Specifically, as increasing magnetic field from zero to 9 T with an initial negative magnetization, the $\sigma_{xy}^{AHE}$ (e.g. at 25 K) increases with the sign change from negative to positive (~ 0.5 T), then decreases with sign change into negative (at ~ 3T), and eventually saturates at large magnetic field (~ 7T).

With increasing temperature, the hump feature becomes weaker and shifts toward lower magnetic field with narrowed field region. The $\sigma_{xy}^{AHE}$ data for other strained samples are also presented in Fig. S3, in which a clearly suppressed hump feature is also observed in tensile-strained sample grown on GSO, but not on any compressive-strained ones, which strongly suggest its direct correlation with epitaxial strain.

With the knowledge of the strain controlled magnetic easy axis, we speculate that the observed hump feature at AHE is related to the magnetization rotation in real space, in which the increasing magnetic field along OOP direction will result in the rotation of magnetization from magnetic easy plane (IP) toward OOP direction for tensile strained SrRuO$_3$ films. To explore the contribution of magnetization rotation to $\sigma_{xy}^{AHE}$, we performed AHE measurement with inclined magnetic field, in which the magnetic field (**B**) is applied with an angle (*θ*) from the film normal (**n**). As shown in Fig. 4b, the hump features are clearly suppressed when slightly tilting angle (15º), and the further increase of *θ* (45º) leads to reappearance of hump features. This observation demonstrates that the hump feature is indeed related to the magnetic rotation. We would like to mention that this point can also exclude the scenario of non-coplanar spin configuration such as magnetic skyrmion (9, 43, 44), in which the hump feature is systematically suppressed and eventually disappears with the tilting above certain angle related with the skyrmion size and sample thickness, due to the destruction of non-coplanar spin structures with inclined magnetic field.

How the rotation of Ru magnetic moment in real space leads to the hump feature observed in SrRuO$_3$ anomalous Hall conductivity can be understood as follows. We first note that the magnetization easy axis of compressive-strained SrRuO$_3$ films is close to OOP. When one sweeps the magnetic field during AHE measurements, the spin-flip results in a well-defined square-like hysteresis loop, as shown in Fig. 2a. By contrast, the magnetic easy axis of tensile-strained

SrRuO₃ films is close to IP direction. With an increasing field from 0 to 9 T along OOP, the spin gradually rotates from its easy axis toward OOP direction due to competition of magnetization anisotropy and Zeeman energy, as illustrated by marks 1 to 4 in Fig. 4a. When magnetic field $B$ is zero (mark 1), magnetization $M$ is along easy axis with a negative remnant magnetization, and a finite anomalous Hall conductivity is measured. With increasing $B$, $M$ rotates into its equivalent easy axis with positive magnetization (mark 2), and the anomalous Hall conductivity changes sign following the change of magnetization. Further increasing $B$ leads to the rotation of $M$ towards the OOP, however, at this point, the anomalous Hall conductivity changes its sign again (mark 3). Eventually at large enough $B$, $M$ is saturated with spin fully aligned along OOP direction, at which the anomalous Hall conductivity also saturates. Clearly, the anomalous Hall conductivity exhibits an interesting non-monotonous evolution with a series of sign changes as magnetic field increases.

To corroborate our speculation about the correlation between the magnetization rotation and the hump feature, we perform first-principles calculations in which we fix the magnetization of SrRuO₃ along different directions in real space. We use $\varphi$ to denote the angle between the magnetization and the out-of-plane sample normal (i.e. the magnetic field direction) (see Fig. 4a). In Fig. 4c, we show the calculated $\sigma_{xy}^{AHE}$ of SrRuO₃ under different tensile strains as a function of $\varphi$. For all strained state, we find that the $\sigma_{xy}^{AHE}$ is very sensitive to the $\varphi$, which should be attributed to the modification of the fermi level as well as electronic band structure with the Zeeman energy. Particularly, at 1% tensile strain (highlighted in red), $\sigma_{xy}^{AHE}$ shows a distinct non-monotonically dependence on $\varphi$ and its sign changes from negative to positive and finally to negative again as $\varphi$ changes from $90°$ (IP) towards $0°$ (OOP). This result qualitatively reproduces the experimental results from mark 1 to 4, as shown in Fig. 4a. Therefore, the characteristic hump feature of $\sigma_{xy}^{AHE}$ observed in tensile-strained SrRuO₃ thin films can be nicely explained by the coupling between the Berry curvature and magnetic rotation.

## Discussion

In summary, we investigate the anomalous Hall effect of SrRuO$_3$ thin films under both compressive and tensile strains. We find that the magnetic easy axis of SrRuO$_3$ is strongly correlated with the epitaxial strain, with the magnetization aligning close to OOP (IP) direction for compressive- (tensile) strained films. Furthermore, the anomalous Hall resistivity (or conductivity) is also highly sensitive to the epitaxial strain, in which negative (positive) anomalous Hall conductivity is observed for films under compressive (tensile) strain at low temperatures. First-principles calculations reveal that epitaxial strain induces an additional crystal field splitting and changes the order of Ru *d* orbital energies, which leads to the sign and magnitude change of anomalous Hall conductivity. Furthermore, we find that the rotation of SrRuO$_3$ magnetization in real space results in a nonmonotonic sign change and a hump feature in the anomalous Hall conductivity, which can be also explained with the evolution of Berry curvature with spin rotation. Our experimental and theoretical results demonstrate that epitaxial strain and the resulting magnetic anisotropy, provide an exciting playground to manipulate Berry curvature and anomalous Hall effect in correlated quantum materials.

## Materials and Methods

**Experimental details.** All samples were prepared by pulsed laser deposition method, with the deposition carried out at 700ºC in an oxygen pressure of 100 mTorr. The laser fluence was set at 2.4 J/cm$^2$ and the repetition rate was 2 Hz. The thicknesses of all strained thin films were measured through x-ray reflectometry, which are between 20 to 30 nm. All transport measurements were performed on a Physical property measurement system (PPMS) equipped with lock-in (LI5640) system.

**DFT calculations.** We perform first-principles calculations, as implemented in Vienna Ab Initio Simulation Package (VASP) (45, 46). We use the generalized gradient approximation with the Perdew-Burke-Ernzerhof parameterization (GGA-PBE) as the exchange-correlation functional (47). We use an energy cutoff of 600 eV and a 10×10×8 Monkhorst-Pack $k$-points to sample the Brillouin zone (48). Spin-orbit coupling (SOC) is included in all calculations. We also consider the correlation effects by performing DFT+$U$+SOC calculations ($U_{Ru}$=2.0 eV) and find that the strain-dependence of anomalous Hall conductivity does not change after $U_{Ru}$ is taken into account (see Fig. S7 in the Supplementary Materials).

Bulk SrRuO$_3$ crystallizes in an orthorhombic structure (*Pnma*, space group No. 62). During the calculations, we first fully relax the crystal structure of SrRuO$_3$ until each force component is smaller than 0.01 eV/Å and pressure on the simulation cell is less than 0.5 kbar. Our calculations find that the orthorhombic structure has lower energy than other crystal structure (cubic and tetragonal) under both compressive and tensile strains (up to 3%). We use the two short lattice constants *a* and *b* to define the pseudocubic lattice constant as $u_0 = \sqrt{ab}$. With respect to the lattice constant of single crystal substrate ($u_0$), we introduce a bi-axial strain $\delta = \frac{u-u_0}{u_0}$. For a given bi-axial strain, we fix the in-plane lattice constant $u_0$, and then optimize the out-of-plane lattice constant *c* and internal atomic coordinates.

**Calculation of Berry curvature and anomalous Hall conductivity.** We use the Wannier90 packages (49, 50) to calculate Berry curvature and anomalous Hall conductivity. All the parameters are obtained from fully relaxed calculations. The strain is defined with respect to the theoretical optimized lattice constant. We use maximally localized Wannier functions (45) to reproduce the DFT-calculated band structure with spin-orbit-coupling in an energy window that includes both Ru $d$ and O $p$ orbitals (see Fig. S4). The intrinsic anomalous Hall conductivity of SrRuO$_3$ is related to its Berry curvature of the underlying electronic structure (12):

$$\sigma_{xy}^{AHE} = -\frac{e^2}{\hbar}\int_{BZ}\frac{d\mathbf{k}}{(2\pi)^3}\Omega^z(\mathbf{k}) \quad (1)$$

where the *z*-component of the Berry curvature is the sum of the band-resolved Berry curvature:

$$\Omega^z(\mathbf{k}) = \sum_n f_n(\mathbf{k})\Omega_n^z(\mathbf{k}) \quad (2)$$

where $f_n$ is the Fermi-Dirac occupancy for the $n^{\text{th}}$ band, and $\Omega_n^z(\mathbf{k})$ is the Berry curvature of the Bloch state $|n\mathbf{k}\rangle$ defined by:

$$\Omega_n^z(\mathbf{k}) = -\text{Im}\left[\langle\partial_{k_x}u_{n\mathbf{k}}|\partial_{k_y}u_{n\mathbf{k}}\rangle - \langle\partial_{k_y}u_{n\mathbf{k}}|\partial_{k_x}u_{n\mathbf{k}}\rangle\right] \quad (3)$$

where $u_{n\mathbf{k}}$ is the periodic part of the Bloch state $|n\mathbf{k}\rangle$. In the actual calculation, we use the Kubo-like formula to compute $\Omega_n^z(\mathbf{k})$:

$$\Omega^z(\mathbf{k}) = -\sum_{mn, m\neq n}(f_n(\mathbf{k}) - f_m(\mathbf{k}))\frac{\text{Im}[v_{nm,x}(\mathbf{k})v_{mn,y}(\mathbf{k})]}{[\epsilon_m(\mathbf{k}) - \epsilon_n(\mathbf{k})]^2} \quad (4)$$

where

$$v_{mn,\alpha}(\mathbf{k}) = \langle\psi_{m\mathbf{k}}|\hat{v}_\alpha|\psi_{n\mathbf{k}}\rangle = \frac{1}{\hbar}\langle u_{m\mathbf{k}}|\partial_{k_\alpha}\hat{H}(\mathbf{k})|u_{n\mathbf{k}}\rangle \quad (5)$$

is a complex velocity and $f_n(\mathbf{k})$ is a Fermi-Dirac occupancy for the Bloch state $|n\mathbf{k}\rangle$. From Eq. (4), If two bands $\epsilon_m(\mathbf{k})$ and $\epsilon_n(\mathbf{k})$ are both filled up or both empty, this pair does not contribute to $\Omega^z$. The largest contribution to $\Omega^z$ comes from an avoided band crossing where spin-orbit interaction opens a small gap and the two bands $m$ and $n$ are such that the $m$-th band is below the Fermi level and the $n$-th band is above the Fermi level.

## Acknowledgement

This work was financially supported by the NSFC (grant Nos. 51872155, 11904196 and 52025024); the Basic Science Center Program of NSFC (grant No. 51788104);

the National Basic Research Program of China (grant No. 2016YFA0301004); the Beijing Natural Science Foundation (Grant No. Z200007); and the Beijing Advanced Innovation Center for Future Chip (ICFC). H.C. is supported by National Natural Science Foundation of China (grant No. 11774236), Open Grant of State Key Laboratory of Low Dimensional Quantum Physics and NYU University Research Challenge Fund.

**Note added:** During the review process, we note that a recent study of $EuCd_2As_2$ (arXiv:2103.09395) (51) also reveals its anomalous Hall effect sensitive to the spin tilting due to the evolution of Berry curvature.

# References and notes

1. M. V. Berry, Quantal Phase-Factors Accompanying Adiabatic Changes. *Proc. R. Soc. Lond. A-Math. Phys. Sci.* **392**, 45-57 (1984).
2. D. Xiao, M. C. Chang, Q. Niu, Berry phase effects on electronic properties. *Rev. Mod. Phys.* **82**, 1959-2007 (2010).
3. D. Vanderbilt, *Berry Phases in Electronic Structure Theory: Electric Polarization, Orbital Magnetization and Topological Insulators* (Cambridge University Press, Cambridge, 2018), DOI: 10.1017/9781316662205.
4. K. von Klitzing, The quantized Hall effect. *Rev. Mod. Phys.* **58**, 519-531 (1986).
5. Y. K. Kato, R. C. Myers, A. C. Gossard, D. D. Awschalom, Observation of the spin Hall effect in semiconductors. *Science* **306**, 1910-1913 (2004).
6. E. N. Adams, E. I. Blount, Energy Bands in the Presence of an External Force Field .2. Anomalous Velocities. *J. Phys. Chem. Solids* **10**, 286-303 (1959).
7. N. Nagaosa, J. Sinova, S. Onoda, A. H. MacDonald, N. P. Ong, Anomalous Hall effect. *Rev. Mod. Phys.* **82**, 1539-1592 (2010).
8. R. Karplus, J. M. Luttinger, Hall Effect in Ferromagnetics. *Phys. Rev.* **95**, 1154-1160 (1954).
9. Y. Taguchi, Y. Oohara, H. Yoshizawa, N. Nagaosa, Y. Tokura, Spin chirality, Berry phase, and anomalous Hall effect in a frustrated ferromagnet. *Science* **291**, 2573-2576 (2001).
10. F. D. M. Haldane, Berry Curvature on the Fermi Surface: Anomalous Hall Effect as a Topological Fermi-Liquid Property. *Phys. Rev. Lett.* **93**, 206602 (2004).
11. Z. Fang *et al.*, The anomalous Hall effect and magnetic monopoles in momentum space. *Science* **302**, 92-95 (2003).
12. Y. G. Yao *et al.*, First principles calculation of anomalous Hall conductivity in ferromagnetic bcc Fe. *Phys. Rev. Lett.* **92**, 4 (2004).
13. S. Itoh *et al.*, Weyl fermions and spin dynamics of metallic ferromagnet $SrRuO_3$. *Nat. Commun.* **7**, 11788 (2016).
14. S. Onoda, N. Sugimoto, N. Nagaosa, Intrinsic versus extrinsic anomalous hall effect in ferromagnets. *Phys. Rev. Lett.* **97**, 4 (2006).
15. J. Ye *et al.*, Berry Phase Theory of the Anomalous Hall Effect: Application to Colossal Magnetoresistance Manganites. *Phys. Rev. Lett.* **83**, 3737-3740 (1999).
16. T. Jungwirth, Q. Niu, A. H. MacDonald, Anomalous Hall Effect in Ferromagnetic Semiconductors. *Phys. Rev. Lett.* **88**, 207208 (2002).
17. C. B. Eom *et al.*, Single-Crystal Epitaxial Thin Films of the Isotropic Metallic Oxides $Sr_{1-x}Ca_xRuO_3$ (0<x<1). *Science* **258**, 1766-1769 (1992).
18. P. B. Allen *et al.*, Transport properties, thermodynamic properties, and electronic structure of $SrRuO_3$. *Phys. Rev. B* **53**, 4393-4398 (1996).
19. G. Koster *et al.*, Structure, physical properties, and applications of $SrRuO_3$ thin films. *Rev. Mod. Phys.* **84**, 253-298 (2012).
20. K. Takiguchi *et al.*, Quantum transport evidence of Weyl fermions in an epitaxial ferromagnetic oxide. *Nat. Commun.* **11**, 4969 (2020).
21. R. Mathieu *et al.*, Scaling of the anomalous Hall effect in $Sr_{1-x}Ca_xRuO_3$. *Phys. Rev. Lett.* **93**, 016602 (2004).
22. R. Mathieu *et al.*, Determination of the intrinsic anomalous Hall effect of $SrRuO_3$. *Phys. Rev. B*

**72**, 064436 (2005).

23. K. J. Choi *et al.*, Phase-Transition Temperatures of Strained Single-Crystal SrRuO₃ Thin Films. *Adv. Mater.* **22**, 759-762 (2010).

24. A. T. Zayak, X. Huang, J. B. Neaton, K. M. Rabe, Manipulating magnetic properties of SrRuO₃ and CaRuO₃ with epitaxial and uniaxial strains. *Phys. Rev. B* **77**, 214410 (2008).

25. A. Vailionis *et al.*, Misfit strain accommodation in epitaxial ABO₃ perovskites: Lattice rotations and lattice modulations. *Phys. Rev. B* **83**, 064101 (2011).

26. B. W. Lee, C. U. Jung, Modification of magnetic properties through the control of growth orientation and epitaxial strain in SrRuO₃ thin films. *Appl. Phys. Lett.* **96**, 102507 (2010).

27. K. Terai, T. Ohnishi, M. Lippmaa, H. Koinuma, M. Kawasaki, Magnetic properties of strain-controlled SrRuO₃ thin films. *Jpn. J. Appl. Phys.* **43**, L227-L229 (2004).

28. D. Kan, R. Aso, H. Kurata, Y. Shimakawa, Epitaxial strain effect in tetragonal SrRuO₃ thin films. *J. Appl. Phys.* **113**, 173912 (2013).

29. A. T. Zayak, X. Huang, J. B. Neaton, K. M. Rabe, Structural, electronic, and magnetic properties of SrRuO₃ under epitaxial strain. *Phys. Rev. B* **74**, 094104 (2006).

30. L. Liu *et al.*, Current-induced magnetization switching in all-oxide heterostructures. *Nat. Nanotechnol.* **14**, 939-944 (2019).

31. W. Lu *et al.*, Strain Engineering of Octahedral Rotations and Physical Properties of SrRuO₃ Films. *Sci. Rep.* **5**, 10245 (2015).

32. Z. Z. Cui *et al.*, Correlation-driven eightfold magnetic anisotropy in a two-dimensional oxide monolayer. *Sci. Adv.* **6**, 8 (2020).

33. J. Matsuno *et al.*, Interface-driven topological Hall effect in SrRuO₃-SrIrO₃ bilayer. *Sci. Adv.* **2**, e1600304 (2016).

34. L. Wang *et al.*, Ferroelectrically tunable magnetic skyrmions in ultrathin oxide heterostructures. *Nat. Mater.* **17**, 1087-1094 (2018).

35. Z. Li *et al.*, Reversible manipulation of the magnetic state in SrRuO₃ through electric-field controlled proton evolution. *Nat. Commun.* **11**, 184 (2020).

36. D. Kan, T. Moriyama, K. Kobayashi, Y. Shimakawa, Alternative to the topological interpretation of the transverse resistivity anomalies in SrRuO₃. *Phys. Rev. B* **98**, 180408 (2018).

37. G. Kimbell *et al.*, Two-channel anomalous Hall effect in SrRuO₃. *Phys. Rev. Mater.* **4**, 054414 (2020).

38. Q. Qin *et al.*, Emergence of Topological Hall Effect in a SrRuO₃ Single Layer. *Adv. Mater.* **31**, 1807008 (2019).

39. P.-C. Wu *et al.*, Thickness dependence of transport behaviors in SrRuO₃/SrTiO₃ superlattices. *Phys. Rev. Mater.* **4**, 014401 (2020).

40. X. Wang, J. R. Yates, I. Souza, D. Vanderbilt, Ab initio calculation of the anomalous Hall conductivity by Wannier interpolation. *Phys. Rev. B* **74**, 195118 (2006).

41. W. Lu, P. Yang, W. D. Song, G. M. Chow, J. S. Chen, Control of oxygen octahedral rotations and physical properties in SrRuO₃ films. *Phys. Rev. B* **88**, 214115 (2013).

42. Anonymous, The shorthand of Ru d orbitals in the bases stands for $|d_{3z^2-r^2}\uparrow\rangle |d_{xz}\uparrow\rangle |d_{yz}\uparrow\rangle |d_{x^2-y^2}\uparrow\rangle |d_{xy}\uparrow\rangle |d_{3z^2-r^2}\downarrow\rangle |d_{xz}\downarrow\rangle |d_{yz}\downarrow\rangle |d_{x^2-y^2}\downarrow\rangle |d_{xy}\downarrow\rangle$.

43. N. Nagaosa, Y. Tokura, Topological properties and dynamics of magnetic skyrmions. *Nat. Nanotechnol.* **8**, 899-911 (2013).

44. T. Yokouchi *et al.*, Stability of two-dimensional skyrmions in thin films of Mn$_{1-x}$Fe$_x$Si investigated

by the topological Hall effect. *Phys. Rev. B* **89,** 064416(2014).

45. G. Kresse, J. Hafner, Ab initio molecular-dynamics simulation of the liquid-metal-amorphous-semiconductor transition in germanium. *Phys. Rev. B* **49**, 14251-14269 (1994).
46. G. Kresse, J. Furthmuller, Efficient iterative schemes for ab initio total-energy calculations using a plane-wave basis set. *Phys. Rev. B* **54**, 11169-11186 (1996).
47. J. P. Perdew, K. Burke, M. Ernzerhof, Generalized Gradient Approximation Made Simple. *Phys. Rev. Lett.* **77**, 3865-3868 (1996).
48. H. J. Monkhorst, J. D. Pack, Special Points for Brillouin-Zone Integrations. *Phys. Rev. B* **13**, 5188-5192 (1976).
49. A. A. Mostofi *et al.*, wannier90: A tool for obtaining maximally-localised Wannier functions. *Comput. Phys. Commun.* **178**, 685-699 (2008).
50. N. Marzari, A. A. Mostofi, J. R. Yates, I. Souza, D. Vanderbilt, Maximally localized Wannier functions: Theory and applications. *Rev. Mod. Phys.* **84**, 1419-1475 (2012).
51. X. Cao *et al.*, Giant nonlinear anomalous Hall effect induced by spin-dependent band structure evolution. *Arxiv:2103.09395* (2021).

# Figures and Captions

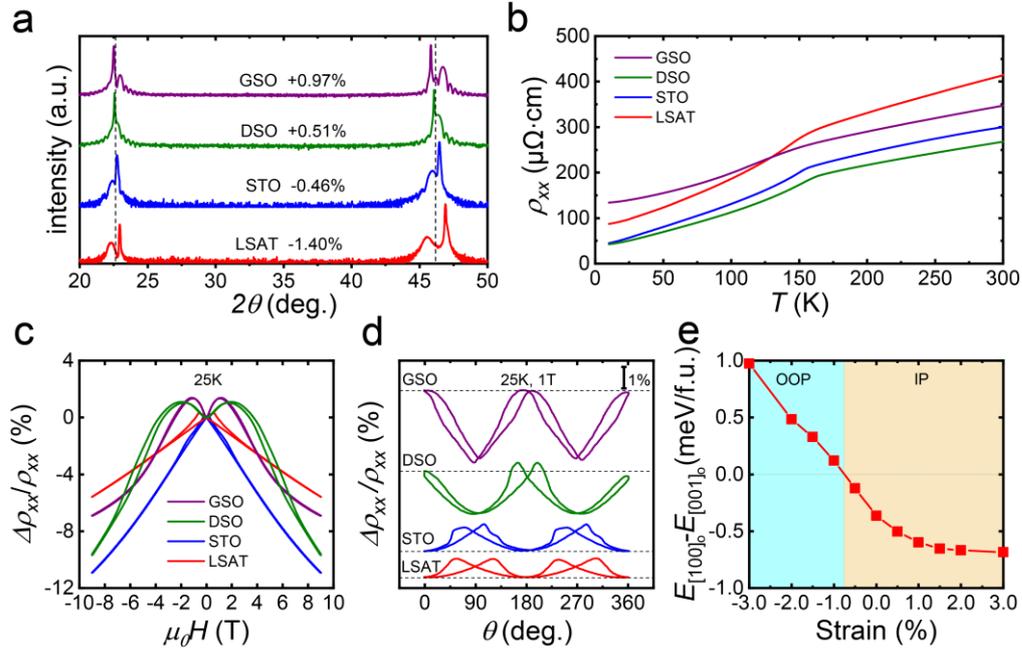

**Fig. 1 | Magnetic anisotropy evolution with epitaxial strain**. **a**, XRD spectra of SrRuO$_3$ thin films grown on various substrates. The dashed lines denote the positions of (001) and (002) peaks for bulk SrRuO$_3$. The strain provided by the substrates is labeled beside each curve, while positive (negative) value indicates the tensile (compressive) strain. **b**, Temperature dependence of longitudinal resistivity $\rho_{xx}$ of different strained SrRuO$_3$ thin films. **c**, Magnetoresistance ($\frac{\Delta\rho_{xx}}{\rho_{xx}} = \rho(H)/\rho(0) - 1$) as a function of magnetic field for different strained SrRuO$_3$ thin films measured at 25 K. The magnetic field was applied along the out-of-plane (film normal) direction. **d**, Angular dependence of anisotropic magnetoresistance (AMR, $\frac{\Delta\rho_{xx}}{\rho_{xx}} = \rho(\theta)/\rho(0) - 1$) for different strained SrRuO$_3$ thin films measured at 25 K with $\mu_0 H = 1$ T. $\theta$ denotes the angle between external magnetic field and the thin film normal direction. The direction of current is along the pseudocubic [110] direction of the thin films, which is perpendicular to magnetic field. The dashed lines indicate AMR = 0. **e**, Strain dependence of the magnetic anisotropy calculated by the energy difference between the cases with the ferromagnetic spins aligned along in-plane [100]$_o$ and out-of-plane [001]$_o$ directions. Here [100]$_o$ and [001]$_o$ are with respect to the orthorhombic notation, which are equivalent to the pseudocubic [110] and [001] directions respectively.

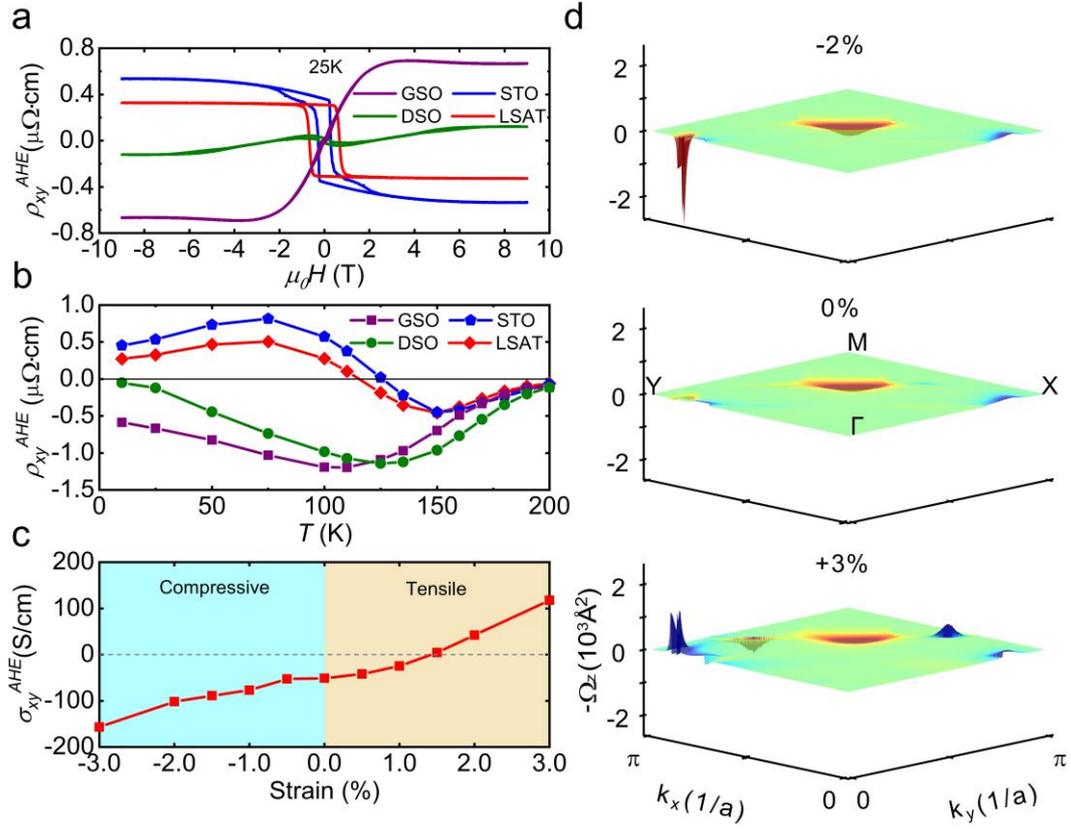

**Fig. 2 | AHE and Berry curvature modulated by epitaxial strain. a**, Anomalous Hall resistivity as a function of magnetic field at 25 K for SrRuO$_3$ thin films with different strains. A linear background due to the conventional Hall effect was subtracted from the Hall signal. **b**, Temperature dependence of saturated anomalous Hall resistivity (taken at -9 T) for different strained SrRuO$_3$ thin films. **c**, Calculated anomalous Hall conductivity as a function of strain. **d**, Itemized calculated Berry curvatures in momentum space for $t_{2g}$ bands as a function of ($k_x$, $k_y$) with $k_z$ being fixed at 0 for different strained SrRuO$_3$ (-2%, 0, 3%).

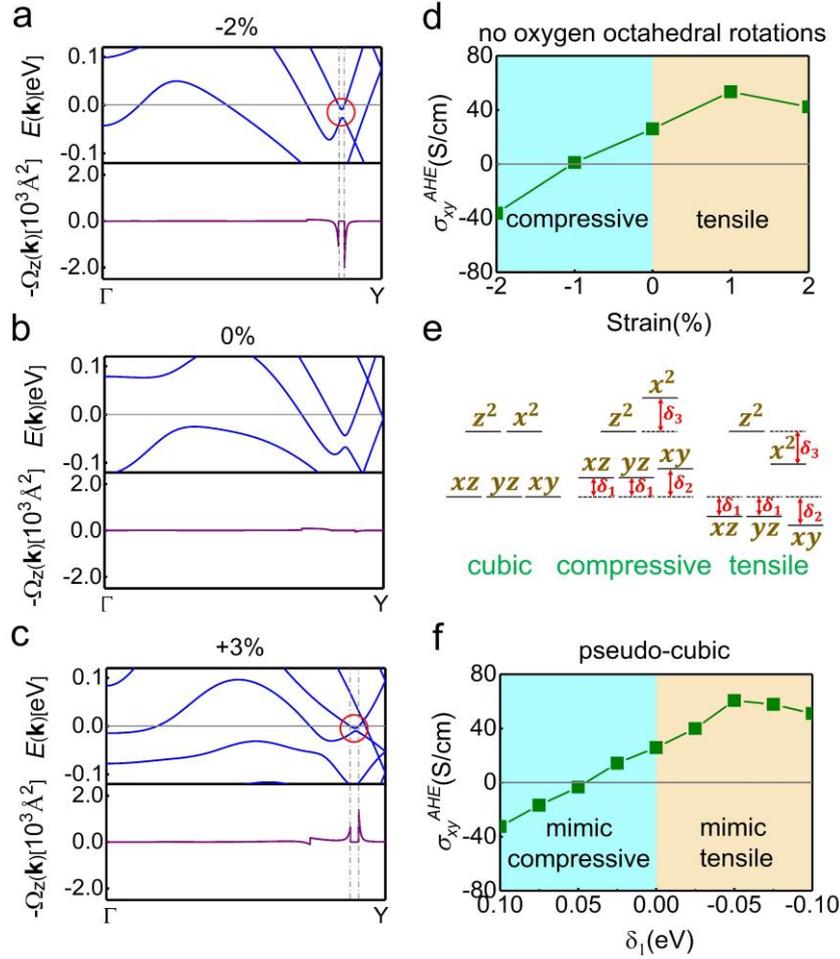

**Fig. 3 | Evolution of electronic structure with epitaxial strain. a-c**, The band structure of SrRuO$_3$ (top panel) and its corresponding Berry curvature (bottom panel) under different strains. **a** 2% compressive strain, **b** no strain and **c** 3% tensile strain. The $x$ axis is a high-symmetry $k$-path from $\Gamma$ to $Y$. **d,** Anomalous Hall conductivity $\sigma_{xy}^{AHE}$ of as a function of epitaxial strain in artificial SrRuO$_3$ with oxygen octahedral rotations manually turned off. **e,** Energy ordering of Ru $d$ orbitals. Left column is pseudo-cubic SrRuO$_3$, in which $d_{xz}$, $d_{yz}$, $d_{xy}$ orbitals are degenerate and $d_{x^2-y^2}$, $d_{3z^2-r^2}$ orbitals are degenerate. Middle column is SrRuO$_3$ under compressive strain, in which $d_{xy}$ orbital has higher energy than $d_{xz}$, $d_{yz}$ orbitals and $d_{x^2-y^2}$ orbital has higher energy than $d_{3z^2-r^2}$. Right column is SrRuO$_3$ under tensile strain, in which $d_{xy}$ orbital has lower energy than $d_{xz}$, $d_{yz}$ orbitals and $d_{x^2-y^2}$ orbital has lower energy than $d_{3z^2-r^2}$. **f,** Anomalous Hall conductivity $\sigma_{xy}^{AHE}$ of pseudo-cubic SrRuO$_3$ as a function of orbital splitting $\delta_1$ ($\delta_2 = 2\,\delta_1$, $\delta_3 = 4\,\delta_1$), which determines the Ru $d$ orbital rearrangement. A positive $\delta_1$ mimics the effects from compressive strain and a negative $\delta_1$ mimics the effects from tensile strain.

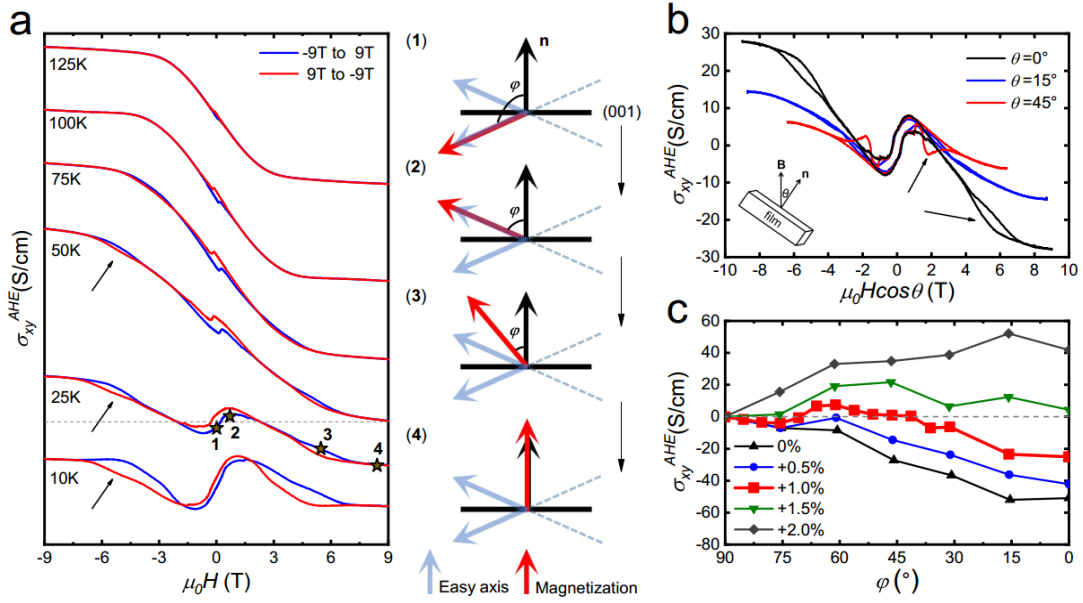

**Fig. 4 | Tuning the AHE and Berry curvature with magnetization rotation. a**, Magnetic field dependence of anomalous Hall conductivity at various temperatures for tensile-strained SrRuO$_3$ thin film grown on DyScO$_3$ substrate. Blue and red lines denote the sweep directions of the magnetic field. Numbered stars denote four representative magnetization (***M***) states. As increasing magnetic field along film normal direction (***n***), ***M*** gradually rotates from easy axis toward ***n*** with an angle ($\varphi$), as shown schematically from state (1) to state (4). **b**, Anomalous Hall conductivity as a function of magnetic field measured at 25 K with varied tilt angles ($\theta$) between the field direction and ***n***. Hump features were observed at 0° and 45° as denoted by arrows. **c**, Calculated anomalous Hall conductivity of SrRuO$_3$ as a function of $\varphi$ under different tensile stains.

# Supplementary Materials of "Manipulate the Berry curvature in SrRuO$_3$ thin films via epitaxial strain"

Di Tian[1#], Zhiwei Liu[2,3#], Shengchun Shen[1#*], Zhuolu Li[1], Yu Zhou[1], Hongquan Liu[3,4], Hanghui Chen[3,4*] and Pu Yu[1,5,6*]

[1]*State Key Laboratory of Low Dimensional Quantum Physics and Department of Physics, Tsinghua University, Beijing, 100084, China*

[2] *Department of Electronics, East China Normal University, Shanghai 200241, China*

[3]*NYU-ECNU Institute of Physics, NYU Shanghai, Shanghai 200122, China*

[4]*Department of Physics, New York University, New York 10012, United States*

[5]*Collaborative Innovation Center of Quantum Matter, Beijing 100084, China*

[6]*RIKEN Center for Emergent Matter Science (CEMS), Wako 351-198, Japan*

*Email: scshen@tsinghua.edu.cn; hanghui.chen@nyu.edu; yupu@tsinghua.edu.cn*

## This PDF file include:

Supplementary Notes
Figs. S1 to S7

## Notes

### 1. Modelling of a pseudo-cubic SrRuO₃ under epitaxial strain

The tight-binding model we use is $H(k) = H_0(k) + H_{soc}$. In the Ru **d** atomic orbital bases $|z^2\uparrow\rangle|xz\uparrow\rangle|yz\uparrow\rangle|x^2\uparrow\rangle|xy\uparrow\rangle|z^2\downarrow\rangle|xz\downarrow\rangle|yz\downarrow\rangle|x^2\downarrow\rangle|xy\downarrow\rangle$ [40], $H_0(k)$ is block diagonal within each spin:

$$H_0(k) = \begin{bmatrix} H_0^\uparrow(k) & 0 \\ 0 & H_0^\downarrow(k) \end{bmatrix} \quad (1)$$

where $H_0^\uparrow(k)$ and $H_0^\uparrow(k)$ are $5\times 5$ matrices and can be determined by using maximally localized Wannier functions. We find that $H_0(k)$ based on the Wannier fitting well reproduces the DFT-calculated band structure (shown in Fig. S5). In these bases $H_{soc} = \lambda \vec{L}\cdot\vec{S}$ has an explicit form, which is shown in Eq. (2)

$$H_{SOC} = \frac{\lambda}{2}\cdot\begin{bmatrix} 0 & 0 & 0 & 0 & 0 & 0 & -\sqrt{3} & i\sqrt{3} & 0 & 0 \\ 0 & 0 & -i & 0 & 0 & \sqrt{3} & 0 & 0 & -1 & i \\ 0 & i & 0 & 0 & 0 & -i\sqrt{3} & 0 & 0 & -i & -1 \\ 0 & 0 & 0 & 0 & -2i & 0 & 1 & i & 0 & 0 \\ 0 & 0 & 0 & 2i & 0 & 0 & -i & 1 & 0 & 0 \\ 0 & \sqrt{3} & i\sqrt{3} & 0 & 0 & 0 & 0 & 0 & 0 & 0 \\ -\sqrt{3} & 0 & 0 & 1 & i & 0 & 0 & i & 0 & 0 \\ -i\sqrt{3} & 0 & 0 & -i & 1 & 0 & -i & 0 & 0 & 0 \\ 0 & -1 & i & 0 & 0 & 0 & 0 & 0 & 0 & 2i \\ 0 & -i & -1 & 0 & 0 & 0 & 0 & 0 & -2i & 0 \end{bmatrix} \quad (2)$$

with row/column labels $z^2\uparrow, xz\uparrow, yz\uparrow, x^2\uparrow, xy\uparrow, z^2\downarrow, xz\downarrow, yz\downarrow, x^2\downarrow, xy\downarrow$.

where $\lambda$ is the atomic spin-orbit-coupling and we find that $\lambda$ = 0.13 meV is reasonable to describe the SOC effects of pseudo-cubic SrRuO₃ (compared to the DFT+SOC results).

To mimic the effects from epitaxial strain, we change the onsite energy of Ru **d** orbitals. We note that we can always add an energy shift on all orbitals, which is counteracted by a shift in the chemical potential (this does not change any physical properties). Therefore, we can assign the following diagonal matrix elements as (assuming that spin up is the majority spin):

$$h_{z^2\uparrow} \rightarrow h_{z^2\uparrow} \quad (3)$$

$$h_{xz\uparrow} \to h_{xz\uparrow} + \delta_1 \tag{4}$$

$$h_{yz\uparrow} \to h_{yz\uparrow} + \delta_1 \tag{5}$$

$$h_{xy\uparrow} \to h_{xy\uparrow} + \delta_2 \tag{6}$$

$$h_{x^2\uparrow} \to h_{x^2\uparrow} + \delta_3 \tag{7}$$

Then we use the revised $H(k) = H_0(k) + H_{\text{soc}}$ and Eq. (1), (2), (4), (5) in Materials and Methods to calculate the $\sigma_{xy}^{AHE}$ as a function of $\delta_1$, $\delta_2$ and $\delta_3$. In the main text, to simply the modelling, we set $\delta_2 = 2\delta_1$ and $\delta_3 = 4\delta_1$. We note that the trend that under compressive strain, $\sigma_{xy}^{AHE}$ becomes negative and under tensile strain, $\sigma_{xy}^{AHE}$ becomes positive, does not depend on the ratio $\frac{\delta_3}{\delta_1} = 4$ and $\frac{\delta_2}{\delta_1} = 2$. We also test other ratios (see Fig. S6) and find that the qualitative trend remains the same as long as the strain-induced orbital splitting is physical: $|\delta_3| > |\delta_2| > |\delta_1|$.

## 2. Epitaxial-strain-induced orbital splitting in transition metal oxides

In transition metal oxides, the metal *d* orbitals hybridize with O *p* orbitals. In the simplest modelling, we study a one-dimensional nonmagnetic transition metal oxide which consists of one metal *d* orbital and one oxygen *p* orbital. Imposing periodic boundary condition leads to a new quantum number *k*.

$$H(k) = \begin{pmatrix} \epsilon_d & t_{pd}(1 - e^{-ik}) \\ t_{pd}(1 - e^{ik}) & \epsilon_p \end{pmatrix} \tag{8}$$

The eigenvalues of Eq. (8) are (expanding in $t_{pd}$):

$$\eta_1(k) = \epsilon_p - \frac{4t_{pd}^2 \sin^2\left(\frac{k}{2}\right)}{\epsilon_d - \epsilon_p} \tag{9}$$

$$\eta_2(k) = \epsilon_d + \frac{4t_{pd}^2 \sin^2\left(\frac{k}{2}\right)}{\epsilon_d - \epsilon_p} \tag{10}$$

Transition metal *d* orbitals are anti-bonding states and therefore the second band $\eta_2(k)$ corresponds to the metal *d* state. The onsite energy of the metal *d* state (the diagonal element of *H* (**R** = 0)) is (in this 1D case):

$$E_d = \int_{-\pi}^{\pi} \frac{dk}{2\pi} \eta_2(k) = \epsilon_d + \frac{2t_{pd}^2}{\epsilon_d - \epsilon_p} \tag{11}$$

In the ferromagnetic case, because magnetization in transition metal oxides mainly resides on metal *d* orbitals, but much less on oxygen *p* orbitals, we will make the following changes to the orbital energies in the ferromagnetic case (assuming that the spin up is the majority spin):

$$\epsilon_d^\uparrow = \epsilon_d, \quad \epsilon_d^\downarrow = \epsilon_d + J, \quad \epsilon_p^\uparrow = \epsilon_p^\downarrow = \epsilon_p \tag{12}$$

Where $J > 0$ is the exchange splitting. Then the Hamiltonian for the ferromagnetic 1D transition metal oxides is:

$$H_\uparrow(k) = H_\uparrow(k) = \begin{pmatrix} \epsilon_d & t_{pd}(1 - e^{-ik}) \\ t_{pd}(1 - e^{ik}) & \epsilon_p \end{pmatrix} \tag{13}$$

$$H_\downarrow(k) = H_\downarrow(k) = \begin{pmatrix} \epsilon_d + J & t_{pd}(1 - e^{-ik}) \\ t_{pd}(1 - e^{ik}) & \epsilon_p \end{pmatrix} \tag{14}$$

The eigenvalues of Eq. (13) and (14) that correspond to the metal *d* orbitals are (expanding in $t_{pd}$):

$$\eta_2^\uparrow(k) = \epsilon_d + \frac{4t_{pd}^2 \sin^2\left(\frac{k}{2}\right)}{\epsilon_d - \epsilon_p} \tag{15}$$

$$\eta_2^\downarrow(k) = \epsilon_d + J + \frac{4t_{pd}^2 \sin^2\left(\frac{k}{2}\right)}{\epsilon_d + J - \epsilon_p} \tag{16}$$

The on-site energy follows:

$$E_d^\uparrow = \epsilon_d + \frac{2t_{pd}^2}{\epsilon_d - \epsilon_p} \tag{17}$$

$$E_d^\downarrow = \epsilon_d + J + \frac{2t_{pd}^2}{\epsilon_d + J - \epsilon_p} \tag{18}$$

Now we impose epitaxial strain which changes the metal-oxygen distance and thus $t_{pd}$. Finally, we obtain (expanding in Δ):

$$\tilde{E}_d^\uparrow = \epsilon_d + \frac{2(t_{pd} + \Delta)^2}{\epsilon_d - \epsilon_p} \simeq \epsilon_d + \frac{2t_{pd}^2}{\epsilon_d - \epsilon_p} + \frac{4t_{pd}}{\epsilon_d - \epsilon_p}\Delta \tag{19}$$

$$\tilde{E}_d^{\downarrow} = \epsilon_d + J + \frac{2(t_{pd} + \Delta)^2}{\epsilon_d + J - \epsilon_p} \simeq \epsilon_d + J + \frac{2t_{pd}^2}{\epsilon_d + J - \epsilon_p} \epsilon_d + \frac{4t_{pd}}{\epsilon_d + J - \epsilon_p}\Delta \quad (20)$$

We find that for the majority spin, the additional crystal field splitting induced by epitaxial strain on Ru *d* orbitals is $\frac{4t_{pd}}{\epsilon_d - \epsilon_p}\Delta$, while for the minority spin, the additional crystal field splitting is $\frac{4t_{pd}}{\epsilon_d + J - \epsilon_p}\Delta$. Since $J > 0$, the additional crystal field splitting on the minority spin is always smaller than that on the majority spin. If $J$ is sufficiently large, the additional crystal field splitting induced by the epitaxial strain on the minority spin can be ignored. This is the approximation we make in our modelling of ferromagnetic pseudo-cubic SrRuO$_3$.

**Supplementary Figures**

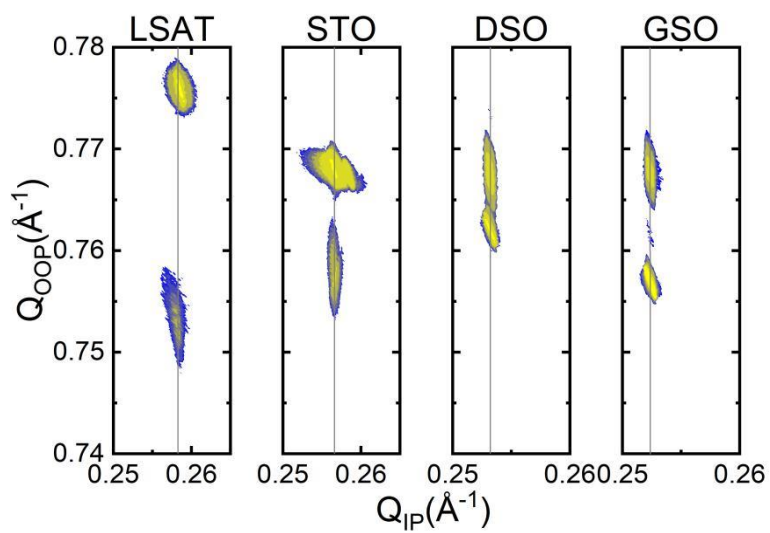

**FIG. S1.** X-ray reciprocal space mapping measurements along the pseudocubic (103) peaks of SrRuO$_3$ films grown on LSAT, STO, DSO and GSO substrates. It is seen that the diffraction peaks of all four films appear at the same in-plane Q vectors as the substrates, indicated by the gray vertical lines, which confirm that all thin films are coherently strained.

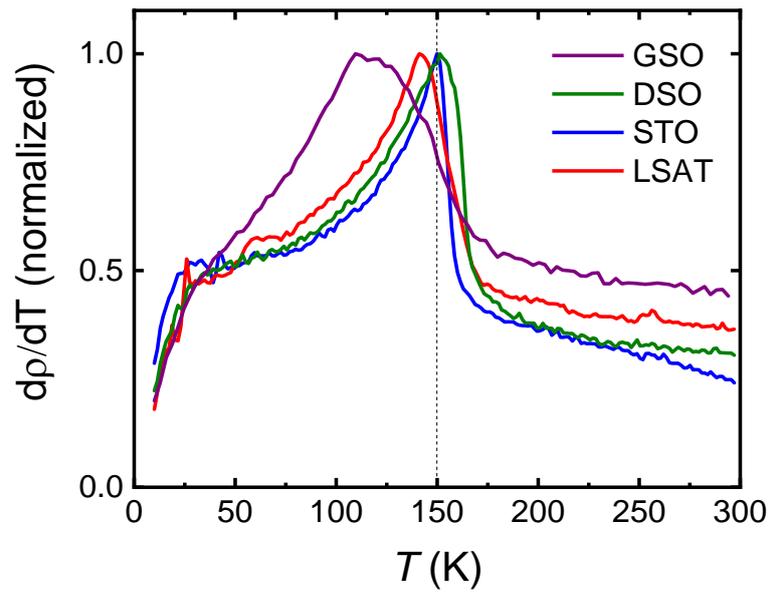

**FIG. S2.** Normalized dρ/dT curves as a function of temperature for SrRuO$_3$ strained films grown on LSAT, STO, DSO and GSO substrates. The peak positions correspond to the curie temperature of the ferromagnetic states.

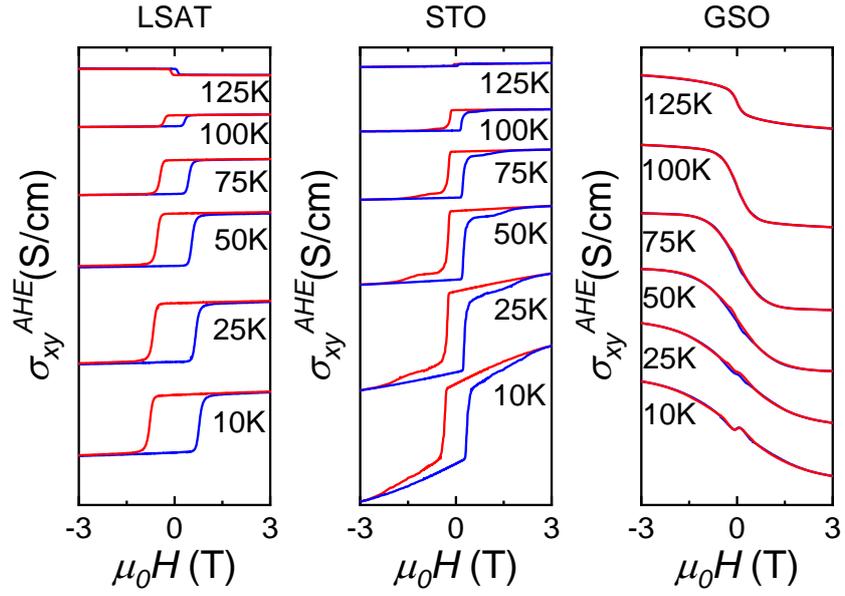

**FIG. S3.** Anomalous Hall conductivity $\sigma_{xy}^{AHE}$ as a function of magnetic field for SrRuO3 films grown on LSAT, STO and GSO substrates, measured from 10K to 125K. The $\sigma_{xy}^{AHE}$ of SRO/STO at 125 K is multiplied by a factor of 10. The field-sweeping direction is from -9T to 9T for blue curves, and from 9T to -9T for red curves.

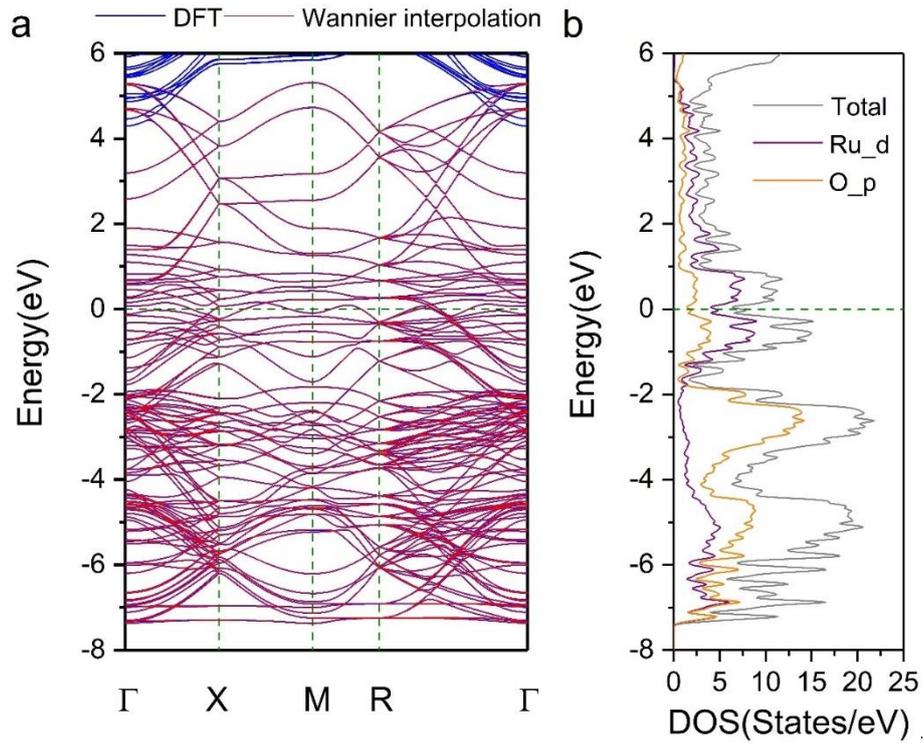

**FIG. S4.** (a) Comparison of the band structures of bulk SrRuO$_3$ obtained from the DFT calculation with spin-orbit coupling and the Wannier interpolation. (b) Total density of state and partial density of state for SrRuO$_3$.

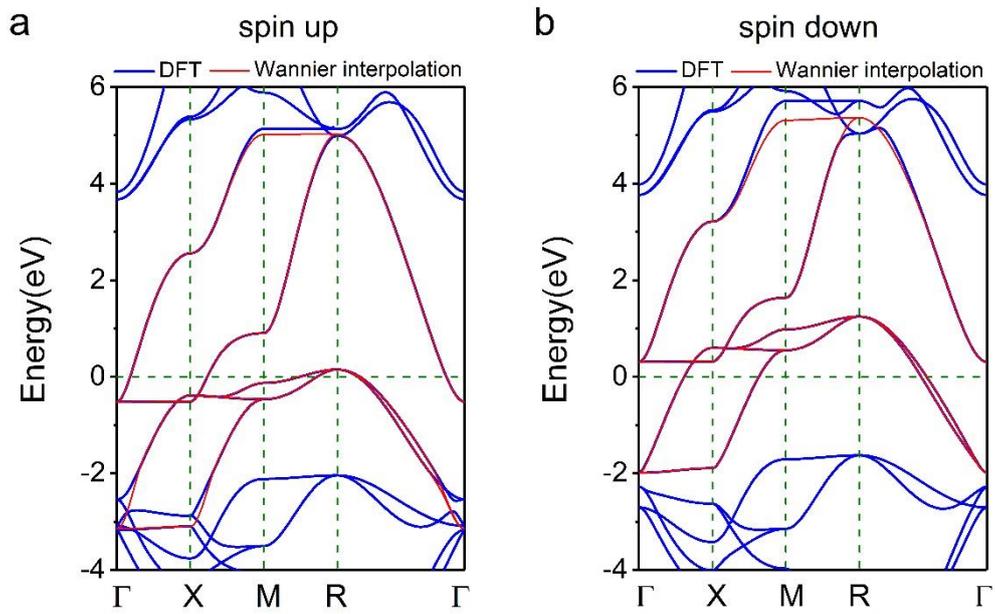

**FIG. S5.** Comparison of the band structures of bulk SrRuO$_3$ without oxygen octahedral rotations obtained from the DFT calculation without spin-orbit coupling and the Wannier interpolation. (a) spin up. (b) spin down.

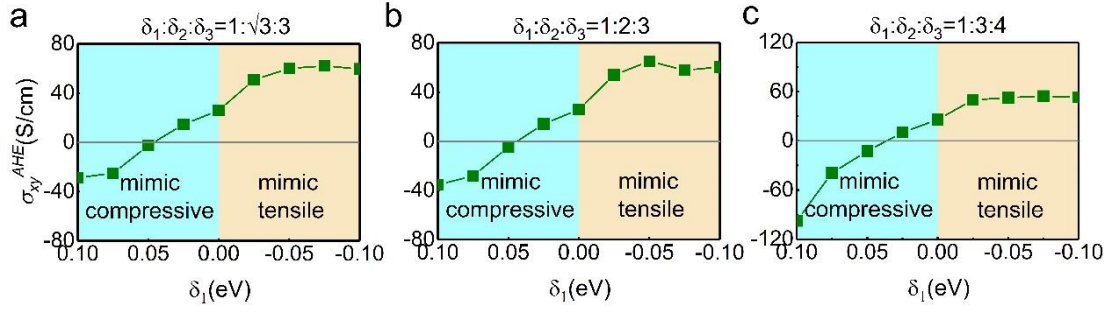

**FIG. S6.** The anomalous Hall conductivity $\sigma_{xy}^{AHE}$ of pseudo-cubic SrRuO$_3$ as a function of orbital splitting $\delta_1$, which determines the Ru *d* orbital ordering. A positive $\delta_1$ mimics the effects from compressive strain and a negative $\delta_1$ mimics the effects from tensile strain. (a) $\delta_1:\delta_2:\delta_3 = 1:\sqrt{3}:3$. (b) $\delta_1:\delta_2:\delta_3 = 1:2:3$. (c) $\delta_1:\delta_2:\delta_3 = 1:3:4$.

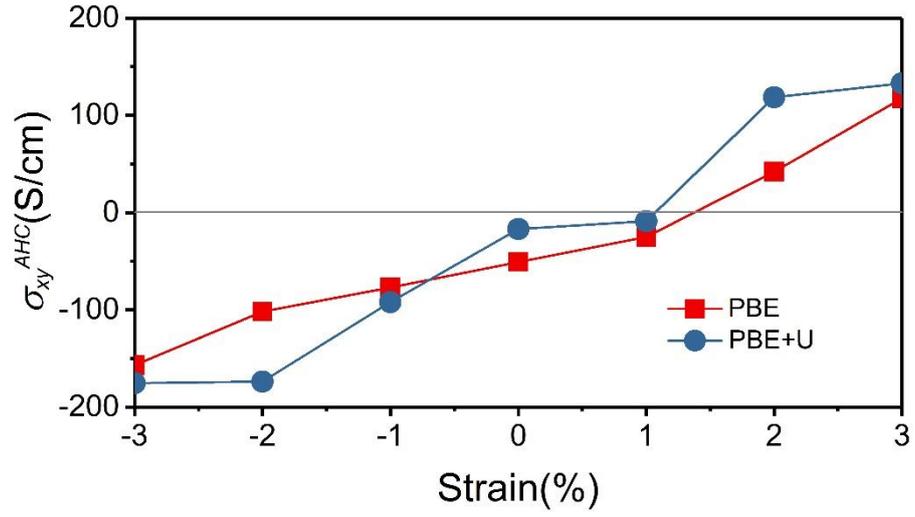

**FIG. S7.** Comparison of the AHC obtained from PBE + SOC calculations and PBE + *U* + SOC calculations with $U_{Ru}$ = 2.0 eV.